%% file: FTZv3.tex
\documentclass[12pt,a4paper]{article}
\usepackage{a4wide}
\usepackage[centertags]{amsmath}
\usepackage{amssymb}
\usepackage{epsfig}
\usepackage{cite}

\renewcommand{\thefootnote}{\fnsymbol{footnote}}

\def\R{{\mathcal R}}
\def\be{\begin{equation}}
\def\ee{\end{equation}}
\def\bea{\begin{eqnarray}}
\def\eea{\end{eqnarray}}

\begin{document}

\thispagestyle{empty}

\begin{flushright}
LTH 748\\
IPPP/07/37\\
DCPT/07/74\\
\end{flushright}
\vskip 2cm

\begin{center}
{\Large \bf Orientifold's Landscape: Non-Factorisable Six-Tori }
\end{center}
\vspace*{5mm} \noindent
\vskip 0.5cm
\centerline{ \large Stefan F\"orste$^{a}$\footnote{
        E-mail address: Stefan.Forste@durham.ac.uk},
      Cristina Timirgaziu$^{b}$\footnote{
        E-mail address: Maria.Timirgaziu@liverpool.ac.uk} and Ivonne
      Zavala$^{a}$\footnote{
        E-mail address: Ivonne.Zavala@durham.ac.uk}}

\vskip 1cm

\centerline{$^a$ \em Institute for Particle Physics Phenomenology (IPPP)}
\centerline{\em South Road, Durham DH1 3LE, United Kingdom}
\vskip0.5cm
\centerline{$^{b}$ \em Department of Mathematical Sciences}
\centerline{\em University of Liverpool,
                Liverpool L69 7ZL}
\vskip2cm

\centerline{\bf Abstract}
\vskip .3cm
We construct type IIA orientifolds on $T^6/{\mathbb Z}_2 \times
{\mathbb Z}_2$ which admit non factorisable
lattices. We describe a method to deal with this kind of
configurations and discuss how the compactification lattice affects
the tadpole cancellation conditions. Moreover, we include D6-branes
which are not parallel to O6-planes.
These branes can give rise to chiral spectra in four dimensions, thus
uncovering a new corner in the landscape of intersecting D-brane model
constructions.
We demonstrate the construction at an explicit example. In general we
argue that obtaining an odd number of families is problematic.

\vskip .3cm

\newpage
%
%
\renewcommand{\thefootnote}{\arabic{footnote}}
\setcounter{footnote}{0}
\section{Introduction}
We are living an exciting era for particle physics, in which, the
first (definite) results from the LHC (Large Hadron Collider)
experiments will put a landmark for this century's particle physics.
It is thus natural that investigation of phenomenological aspects of
every BSM (Beyond Standard Model) theory be in a flourishing season at
this time.

In particular, string theory, as one of the best candidates for a
unifying theory, has seen a great deal of development, both in its
phenomenological as well as cosmological aspects in recent
times.
From the phenomenological point of view, D-brane models
\cite{Sagnotti:1987tw,Horava:1989vt,Govaerts:1988md,Bern:1989zr,Gimon:1996rq}
are among the most interesting set-ups that have been widely  studied
in both aspects lately.

More concretely, intersecting D-brane models in  toroidal
orientifold compactifications, represent a simple yet rich
set-up. Particle physics models, with spectra close to that of
the Standard Model, both in non-supersymmetric
\cite{Blumenhagen:2000wh,imr,Blumenhagen:2001te,Angelantonj:2000hi,Aldazabal:2000cn,Forste:2001gb,Bailin:2001ie,koko,Cremades:2002dh,Ellis:2002ci,Abel:2003yx}
as well as supersymmetric
\cite{Cvetic:2001tj,Blumenhagen:2002gw,Cvetic:2002pj,cll,Honecker:2003vq,Larosa:2003mz,Dudas:2005jx,Blumenhagen:2005tn,Bailin:2006zf,Font:2006na}
configurations, can be constructed (for reviews and more
references see \cite{Angelantonj:2002ct}).

In this context, most models considered so far in the {\em
  landscape}\footnote{By landscape here we mean the whole set of
  possibilities that can arise within string theory constructions. In
  particular, we are interested in orientifold constructions in
  type IIA theories.  } of possibilities, have been restricted to
factorisable tori \cite{Gmeiner:2005vz}\footnote{For statistical
  investigations of more abstract CFT orientifolds see
  \cite{Dijkstra:2004ym}.}.
That is, toroidal compactifications where the six dimensional internal
  manifold, $T^6$, can be factorised as the product of three two-tori
  $T^6 = T^2\times T^2\times T^2 $.
However, it is very natural to ask whether more generic toroidal
  orientifold  compactifications can be constructed.
A first step in this direction has already been taken in heterotic
  theory
  \cite{Dixon:1985jw,Erler:1992ki,heterotic1,heterotic2,heterotic3,Takahashi:2007qc},
  where orbifold
  models, which admit more complicated lattices, i.e. non factorisable
  lattices, were analysed. (An interesting observation is that some
  of the spectra can be obtained on factorisable tori with the notion
  of generalised torsion \cite{Ploger:2007iq}.)
Also, in the context of Type IIA theory, non-chiral models of
  orientifolds for supersymmetric ${\mathbb Z}_N$ orbifolds, in
  non-factorisable tori were constructed in
  \cite{Blumenhagen:2004di}. Using a different language, in
  \cite{Sagnotti:1987tw}, 
  a six dimensional $\mathbb Z_2$ orientifold of an SO(8) lattice in
  Type IIB theory was presented. 

In  the case of factorisable orientifold models,  the phenomenological
requirement of getting an odd number of families (three) puts a strong
constraint on the geometry of the factorisable torus
\cite{Blumenhagen:2000wh}. As it has been shown
\cite{Blumenhagen:2000wh}, it is necessary to introduce, besides
untilted $T^2$ tori, or type {\bf A} two dimensional lattices
\cite{Blumenhagen:1999md}, also tilted tori, or type {\bf B} lattices.

Pure orientifold (that is, no orbifold action performed)
non-supersymmetric models,  with spectra very close to that of the
Standard Model,  were first constructed in \cite{imr} along these
lines, where one of the three two-tori was tilted. One problem of
these non-supersymmetric models, arises from stability issues, due to
the presence of NSNS tadpoles and tachyons.
A supersymmetric version,  in the factorisable orbifold $T^6/{\mathbb
  Z_2}\times {\mathbb Z_2}$, appeared in \cite{Cvetic:2001tj}, where
again, only one of the three two-tori was tilted\footnote{It was
shown in \cite{cll} that, for phenomenological purposes, introducing
two or three tilted tori in these  supersymmetric models,  provides
no solution. Therefore,  one has to stick to single tilted tori set-ups.}.
Such supersymmetric constructions are more under control from
the point of view of stability, but contain typically chiral exotic
states.

In this note, we take a diversion from the usual factorisable path and
explore orientifolds of
$T^6/{\mathbb Z_2}\times {\mathbb Z_2}$ which admit non-factorisable
lattices. In particular, we show that such generalisations are easy
to deal with. Moreover, we incorporate  parallel as well as
non-parallel D6-branes,  which can then give rise to  four dimensional
chiral spectra.
We  discuss how the tadpole cancellation conditions arise in this more
general models and construct an explicit example. Although for
phenomenological (and stability) reasons, we focus mainly on four
dimensional ${\mathcal N}=1$
 models, we also comment on  how non-supersymmetric set-ups can be
 implemented, which can have interesting phenomenology,
 in spite of suffering from possible instabilities due to the lack of
 supersymmetry.

We start in the next section by reviewing the $T^6/{\mathbb Z_2}\times
{\mathbb Z_2}$ orientifold construction in the factorisable case, and
fix our notation.
We then turn in section 3  to  non factorisable models. We consider
explicitly  the SO(12) root lattice, as an illustrative example  of our
method of dealing with non factorisable tori. We show how to compute
the tadpole cancellation conditions for parallel (to the orientifolds)
branes. We then  introduce non parallel branes, which are invariant
under  the orbifold action, and which preserve supersymmetry
(although this is not strictly required). We present an explicit
model as an example of these constructions.

In section 4, we elaborate on how more general lattices, i.e.~non
factorisable, put severe constraints on the wrapping numbers.
This fact gets then reflected in the intersection numbers, which
ultimately are directly connected to the number of families in the
models. We show that these constraints give rise generically to an even
number of families, irrespective of supersymmetry requirement, if one
sticks to orbifold invariant D-branes.
We then consider the possibility of having  non-invariant branes, and
show that whereas supersymmetric models are clearly excluded for
giving  only even number of generations, non supersymmetric models
could still be constructed with spectra close to that of the Standard
Model. However, as in the factorisable case, the stability of such
constructions is not guaranteed.
Finally we present some comments and conclusions in the last section.


\section{Recap of the factorisable \boldmath{${\mathbb Z}_2 \times
    {\mathbb Z}_2$}   orientifold} \label{recap}

In this section we rephrase the derivation of known results for
${\mathbb Z}_2 \times {\mathbb Z}_2$ orientifolds of factorisable
six-tori \cite{Berkooz:1996dw,Forste:2000hx}. Specifically, we
perform orientifolds of type IIA strings as in
\cite{Forste:2000hx}. These are related to the original type IIB
model \cite{Berkooz:1996dw} and its generalisations with discrete
B-fields \cite{Kakushadze:1998eg} by T-duality. The effects of
the antisymmetric background tensor in pure orientifold
compactifications were first described in \cite{Bianchi:1991eu}.
In \cite{Kakushadze:1998bw} these results
 have been extended to the case of
$T^4/{\mathbb Z_N}$ orientifolds, while in
\cite{Angelantonj:1999jh} a more detailed analysis was performed
and carried further to four dimensional compactifications as
well. Finally, in \cite{Angelantonj:1999xf} the connection with
type IIA orientifolds through T-dualities was discussed.

\begin{figure}[h!]
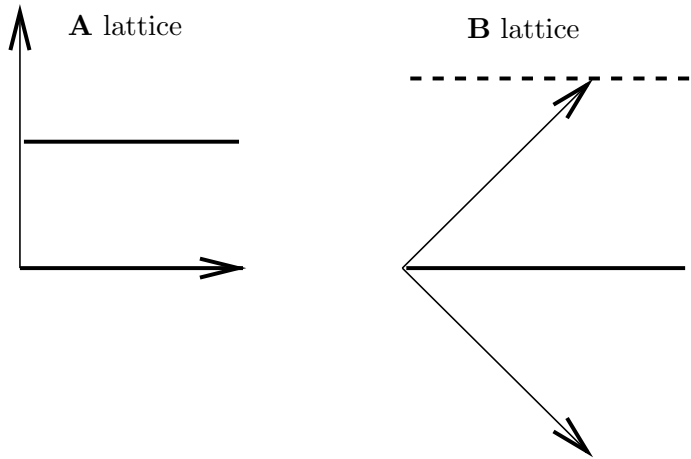

\centering
\begin{center}\input 2dfix.pstex_t \end{center}
\caption{\it The two dimensional compactification lattices: Lattice
  vectors are drawn with thinner lines than fixed lines. The dashed
  line in the right figure differs from the thick solid line by a
  lattice shift. Note, that for the {\bf A} lattice the lengths of the
  two basis vectors can be different.}
\label{fig:2dlat}
\end{figure}

When we talk about
factorisable six-tori we mean that a decomposition into the product of
three two-tori is respected by orbifold and orientifold
actions, i.e.\ each factor is mapped onto itself. It is important to
note that the notion of factorisable (or
non factorisable) makes sense only in combination with the orbifold and
orientifold actions. Further one has to specify the dimensionality of
the factors (two in our case).
Each of the two-tori can be viewed as a compactification of
a complex plane which we parameterise by complex coordinates $z^i$,
$i=1,2,3$. We specify the orbifold and orientifold actions by
their action on these coordinates. Calling the ${\mathbb Z}_2 \times
{\mathbb Z}_2$ generators $\theta$ and $\omega$ we explicitly assign
\begin{equation}
\theta\, z^i = e^{2 \pi \mbox{\scriptsize i}\, v_i}\,\, z^i \,\,\, ,
  \,\,\, \omega \, z^i =
  e^{2\pi \mbox{\scriptsize i}\, w_i}\,\, z^i ,
\label{eq:orbaction}
\end{equation}
with
\begin{equation}
\vec{v} = \left( \frac{1}{2}, -\frac{1}{2}, 0 \right) \,\,\, , \,\,\,
\vec{w} = \left( 0, \frac{1}{2}, -\frac{1}{2}\right) .
\end{equation}
Further, the orientifold element $\Omega\R$ acts as world sheet parity
inversion $\Omega$ together with complex conjugation
on the coordinates
\begin{equation}
\R\, z^i = \overline{z}^i \, , \,\,\, i = 1, 2, 3.
\end{equation}
To obtain a product of three two-tori we equip each of the three
complex planes with a two dimensional compactification lattice. That
lattice has to be invariant under orbifold and orientifold
actions. The two possible choices are \cite{Blumenhagen:1999md}
\begin{itemize}
\item the {\bf A} lattice is spanned by $(1,0)$  and $(0,1)$ ,
\item the {\bf B} lattice is spanned by $(1,1)$ and $(1,-1)$.
\end{itemize}
Both these compactifications can be decomposed into a product of two
circles. Only for the {\bf A} lattice the $\R$-action respects that
decomposition. There are two $\R$-fixed lines for the {\bf A} lattice
and one for the {\bf B} lattice. The situation is depicted in figure
\ref{fig:2dlat}.
In analogy to our definition of factorisable six-tori, 
the {\bf A} lattice can be viewed as factorisable into a product of
two circles whereas the {\bf B} lattice cannot. It is because of this
analogy that the rederivation to be discussed in this section is
useful. We
use a language which makes a generalisation to non factorisable
six-tori (into two-tori) straightforward.

Here, we focus on the case that we introduce D-branes which
are parallel to O-planes in order to cancel their RR charges.
Each {\bf A} lattice contains twice as many O-planes as the {\bf
  B} lattice. Therefore each time one replaces an {\bf A} lattice by a
{\bf B} lattice the number of D-branes is reduced by a factor 1/2.
When we perform the calculation in the string loop channel the
number of O-planes does not enter directly. In order to compute
the total RR charge one performs a modular transformation to the
string tree channel. The zero mode part of the modular
transformation consists of Poisson resummations for windings and
momenta. This introduces factors depending on the
compactification lattice and hence the number of O-planes. In the
following we  use this method to obtain the anticipated factors
of 1/2 for {\bf B} lattices from a loop channel calculation.

First, focus on the Klein bottle and in particular on the
contribution with an $\Omega {\cal R}$ insertion (other insertions
like $\Omega {\cal R}\theta$  are straightforward
modifications). Windings and momenta need to be invariant
under the $\Omega {\cal R}$ insertion. This means that momenta are on
the ${\cal R}$ invariant sublattice of the dual compactification
lattice, but this is the dual of an ${\cal R}$ projected
lattice\footnote{The ${\cal R}$ projected lattice is obtained by
  acting with the operator $\left(1+{\cal R}\right) /2$ on the
  lattice.}, which
we call $\Lambda_{{\cal R}, \perp} ^\star$.\footnote{
This can be seen by modifying the argumentation in Appendix A of
\cite{Narain:1986qm} (see also \cite{Erler:1992ki,heterotic3}). First,
we show that every element of $\Lambda_{{\cal R}, \perp} ^\star$ is in
the invariant sublattice of the dual lattice $\Lambda^\star$. The
projected lattice 
$\Lambda_{{\cal R}, \perp}$ is three dimensional and so is its
dual. Any element of $\Lambda_{{\cal R}, \perp} ^\star$ can be written
as $\left( x,0,y,0,z,0\right)$, and has
integer valued scalar products with every element of  $\Lambda_{{\cal
    R}, \perp}$. Because the 2$^{nd}$, 4$^{th}$ and 6$^{th}$ component
do not enter, the scalar product with any element of $\Lambda$ is
integer valued. Hence, every vector in  $\Lambda_{{\cal R}, \perp}
^\star$  lies also in $\Lambda^\star$, and is obviously in its invariant
sublattice. It remains to show that also every vector in the invariant
sublattice of $\Lambda^\star$ is in $\Lambda_{{\cal R}, \perp}
^\star$. Any ${\cal R}$-invariant vector $v \in \Lambda^\star$
satisfies
$$ \left( \frac{ 1 + {\cal R}}{2} v, q\right) \in {\mathbb Z} $$
for every $q \in \Lambda$. Since ${\cal R}$ is symmetric it follows
that $v$ has integer valued scalar product with every vector of
$\Lambda_{{\cal R},\perp}$, and hence is in $\Lambda_{{\cal R},\perp}^\star$.}
Since $\Omega$ gives an
additional sign for
windings these are quantised on a $-{\cal R}$ invariant lattice which
we call $\Lambda_{-{\cal R}, inv}$. In computing the RR tadpoles one
performs a transformation from the string loop channel to the tree
channel. For the contributions due to windings and momenta, this
implies two Poisson resummations. For each of the $T^2$ factors we obtain a
factor (see e.g.\ footnote 17 in \cite{Erler:1992ki})
\begin{equation}
\frac{\mbox{vol}\left(    \Lambda_{{\cal R}, \perp}\right)}{\mbox{vol}
  \left( \Lambda_{-{\cal R}, inv}\right)} .
\end{equation}
For the {\bf A} lattice this factor is just one whereas for the {\bf
  B} lattice it
is one half (the invariant lattice is generated by (2,0) in this
case). So, finally we obtain the RR tadpole contribution from
the Klein bottle as (up to some overall constants which we  also
suppress in the M\"obius strip and Cylinder diagrams)
\begin{equation}
\mbox{\bf KB:}\,\,\, 2^{-\sum_{i=1}^3\delta_{Bi}} 32^2 ,
\end{equation}
where $\delta_{Bi}$ is one (zero) if the $i$th plane is compactified
on a {\bf B} ({\bf A}) lattice.

The simplest way of canceling the RR charges of the O-planes is to
add D-branes parallel to the O-planes. Let us focus on the set of
D6-branes extended along the real axes of the compact space. The
open string momenta take values on the dual of the ${\cal R}$ invariant
lattice $\Lambda_{{\cal R},inv} ^\star$, whereas windings are
transverse to the brane. One subtlety is that the open string has to
end only on the same brane but not at the same point. Therefore
windings take values on the $-{\cal R}$ projected lattice
$\Lambda_{-{\cal R}, \perp}$. For the contribution to the cylinder
amplitude
with no further insertions into the trace Poisson resummations yield a
factor of
\begin{equation}
\frac{\mbox{vol}\left(    \Lambda_{{\cal R}, inv}\right)}{\mbox{vol}
  \left(\Lambda_{-{\cal R}, \perp }\right)} .
\end{equation}
The relevant contribution to the cylinder amplitude reads
\begin{equation}
\mbox{\bf C:}\,\,\, 2^{\sum_{i=1}^3\delta_{Bi}} N^2 ,
\end{equation}
where $N$ is the number of D-branes.

For the M\"obius strip we focus again on the contribution with the
$\Omega {\cal R}$ insertion. The momentum modes are just the same as
in the cylinder case, i.e.\ on
$\Lambda_{{\cal R},inv} ^\star$. However, because $\Omega$ swaps the
two ends of the open string it has to end in the same point on the
D-brane.  The winding modes need to be
invariant under  $\Omega {\cal R}$, i.e.\ the projected lattice has to
be replaced by the invariant one $\Lambda_{-{\cal R },inv} $. The two
factors appearing due to Poisson resummation cancel as well for the
{\bf A} as the {\bf B} lattice. The M\"obius strip contribution to the
RR tadpole reads
\begin{equation}
- 2\cdot 32\cdot N .
\end{equation}
Adding up the contributions from Klein bottle, M\"obius strip and
Cylinder one obtains the tadpole cancellation condition
\begin{equation}
2^{\sum_{i=1}^3\delta_{Bi}}\left( 32 \cdot
2^{-\sum_{i=1}^3\delta_{Bi}} - N\right)^2 = 0,
\end{equation}
which coincides with the result of \cite{Forste:2000hx}.

For later use we note that we could have just considered the {\bf B}
lattice and obtained the {\bf A} lattice result by a modified
orientifold action.
Instead of changing a lattice from {\bf B} to {\bf A} we can combine
the complex conjugation with a multiplication with  i:
\begin{equation}\label{eq:BtoA}
\mbox{Instead of {\bf B} $\to$ {\bf A} modify ${\cal R}$:}\,\,\, z^i \to
\mbox{i} \overline{z}^i,
\end{equation}
where $i$ labels the plane in which we want to replace the
compactification lattice.
Focusing on the corresponding  $T^2$ factor, we find that the ${\cal
  R}$ projected
lattice is generated by $(1,1)$ whereas the $-{\cal R}$ invariant
lattice is generated by $(1,-1)$. The volumes of the two lattices are
the same and we obtain, as expected, the same result as if we had
considered an {\bf A} compactification. For non factorisable $T^6$, it
is convenient, in some cases, to fix the lattice
and consider different orientifold actions, instead.


\section{Non factorisable lattices: SO(12)}

As a concrete example of a non factorisable $T^6$, and to exemplify our
method,  we study a compactification on an SO(12) root lattice with basis vectors:
\begin{equation}
\begin{aligned}
e_1 & = & \left( 1 , -1 , 0,0,0,0\right) ,  \\
e_2 & = & \left( 0, 1, -1, 0,0,0\right) ,  \\
e_3 & = & \left( 0 ,0 , 1, -1,0,0\right) , \\
e_4 & = & \left( 0, 0, 0, 1,-1,0\right) ,  \\
e_5 & = & \left( 0,0,0,0,1,-1\right) ,  \\
e_6 & = & \left( 0,0,0,0,1,1\right) .
\end{aligned}
\label{eq:so12roots}
\end{equation}
Oscillator contributions to amplitudes do not depend on the
compactification lattice. The discussion of the previous section
can be carried over to non factorisable $T^6$ in a straightforward
manner.

\subsection{Parallel O-planes and D-branes}

The orientifold group acts as before on the coordinates in which the
lattice vectors (\ref{eq:so12roots}) are given. For the element
containing world sheet
parity reversal we take as before $\Omega {\cal R}$ with ${\cal R}$
acting as complex conjugation on the three planes. Then the ${\cal R}$
projected lattice is generated by
\begin{equation}
\Lambda_{{\cal R}, \perp}:\,\,\, \begin{array}{c} \left(
  1,0,0,0,0,0\right) ,\\
\left( 0,0,1,0,0,0\right) , \\
\left( 0,0,0,0,1,0\right)
\end{array}
\end{equation}
and has volume one.
The $-{\cal R}$ invariant lattice is given by SO(6) simple roots
\begin{equation}
\Lambda_{-{\cal R}, inv}:\,\,\, \begin{array}{c} \left(
  0,1,0,-1,0,0\right) ,\\
\left( 0,0,0,1,0,-1\right) , \\
\left( 0,0,0,1,0,1\right) .
\end{array}
\end{equation}
The volume of the fundamental cell is the square root of the
determinant of the SO(6)
Cartan matrix which is two. Hence there is a factor of one half
appearing in the Klein bottle amplitude as compared to the
factorisable {\bf AAA} compactification. In a very similar way we
obtain a factor of two in front of the Cylinder amplitude and conclude
that the tadpole cancellation condition is
\begin{equation}
\left( 16 - N\right) ^2 = 0.
\end{equation}
This result actually holds for all types of D6-branes parallel to
O6-planes (e.g.\ the consideration of $\Omega {\cal R} \theta$, $\Omega {\cal R} \omega$ and $\Omega {\cal R} \theta\omega$
orientifold
fixed planes is completely analogous to the one we carried out, here).
So, we obtain the same condition as for e.g.\ the {\bf BAA}
factorisable compactification. Now, however, there is no
distinguished complex plane. Actually, we can count the number of
O6-planes and confirm that the above result is consistent with the
general rule that an O6-plane carries four D6-brane charges. Four
inequivalent $\Omega {\cal R}$ fixed planes\footnote{The treatment of
  the other O6-planes is completely analogous.} are given by ($z^i =
x^i + \mbox{i} y^i$)
\begin{eqnarray}
\left( x^1, 0, x^2 , 0, x^3,0\right) &, &   \left( x^1, \frac{1}{2},
x^2 , \frac{1}{2}, x^3,0\right) , \nonumber \\
 \left( x^1, \frac{1}{2}, x^2 ,0 , x^3,\frac{1}{2}\right) &, &
 \left( x^1,0, x^2 ,\frac{1}{2} , x^3,\frac{1}{2}\right) .
\label{eq:fourfix}
\end{eqnarray}
Note that e.g.\ the fixed plane $\left( x,1,y,0,z,0\right)$ is related
by a lattice shift $\left( 1,-1,0,0,0,0\right)$ to the first plane in
(\ref{eq:fourfix}).

If we just focus on two dimensional sublattices the SO(12) lattice
looks similar to a {\bf BBB} lattice. In order to point out the
difference to the factorisable case, we call our orientifold {\bf CCC}
model. As described at the end of the previous section we can obtain
different models from the same compactification lattice by replacing
the orientifold action according to (\ref{eq:BtoA}). For each plane
where such a change is performed we replace a {\bf C} by a {\bf D} in
the name of the model.

For instance, we want to change the
orientifold action such that we obtain a {\bf DCC} model. We replace
the ${\cal R}$ action as follows:
\begin{equation}
{\cal R}: \,\,\, z^1 \to \mbox{i}\overline{z}^1,\,\,\, z^i \to
\overline{z}^i, \,\,\, i=2,3 .
\label{eq:ABB}
\end{equation}
The ${\cal R}$ projected lattice is generated by
\begin{equation}
\Lambda_{{\cal R}, \perp}:\,\,\, \begin{array}{c} \left(
  \frac{1}{2},\frac{1}{2},0,0,0,0\right) ,\\
\left( 0,0,1,0,0,0\right) , \\
\left( 0,0,0,0,1,0\right)
\end{array}
\end{equation}
where the first lattice vector can be obtained by acting with $\frac{1
  +{\cal R}}{2}$ on e.g.\ $(1,0,0,0,0,1)$. The volume of the
  fundamental cell is $1/\sqrt{2}$. The $-{\cal R}$ invariant lattice
  is generated by
\begin{equation}
\Lambda_{-{\cal R}, inv}:\,\,\, \begin{array}{c} \left(
  1,-1,0,0,0,0\right) ,\\
\left( 0,0,0,1,0,-1\right) , \\
\left( 0,0,0,1,0,1\right)
\end{array}
\end{equation}
and has volume $\sqrt{8}$. Carrying out a similar consideration for
the lattices appearing in the cylinder amplitude one obtains the
tadpole cancellation condition
\begin{equation}
\left( 8 - N\right)^2 = 0.
\end{equation}
So, we expect to have two O6-planes per orientifold group element. The
two planes for the $\Omega {\cal R}$ element are
\begin{equation}
\left( x^1, x^1, x^2, 0, x^3, 0\right) \,\,\, , \,\,\, \left( x^1,
x^1, x^2, \frac{1}{2}, x^3, \frac{1}{2}\right) .
\end{equation}
Note that  $\left( x^1, x^1 +1, x^2, 0, x^3, 0\right)$
is at the same position as the first plane as can be seen by adding
the SO(12) lattice vector $(0,-1,1,0,0,0)$. Note also that $\left(
x^1, x^1 + \frac{1}{2}, y, \frac{1}{2}, z, 0 \right)$ is not a fixed
plane under the modified action (\ref{eq:ABB}).

For the {\bf DDC} model we take ${\cal R}$ to act as
\begin{equation}
{\cal R}: \,\,\, z^i \to \mbox{i}\overline{z}^i,\,\,\, z^3 \to
\overline{z}^3, \,\,\, i=1,2 .
\label{eq:AAB}
\end{equation}
In this case the ${\cal R}$ projected lattice is generated by
\begin{equation}
\Lambda_{{\cal R}, \perp}:\,\,\, \begin{array}{c} \left(
  \frac{1}{2},\frac{1}{2},0,0,0,0\right) ,\\
\left( 0,0,\frac{1}{2},\frac{1}{2},0,0\right) , \\
\left( 0,0,0,0,1,0\right)
\end{array}
\end{equation}
and its volume is one half. The $-{\cal R}$ invariant lattice has the
following basis
\begin{equation}
\Lambda_{-{\cal R}, inv}:\,\,\, \begin{array}{c} \left(
  1,-1,0,0,0,0\right) ,\\
\left( 0,0,1,-1,0,0\right) , \\
\left( 0,0,0,0,0,2\right) .
\end{array}
\end{equation}
The fundamental cell has volume four. This and similar considerations
lead to the tadpole cancellation condition
\begin{equation}
\left( 4 - N\right)^2 = 0.
\end{equation}
Now there is just one $\Omega {\cal R}$ fixed plane at
\begin{equation}
\left( x^1, x^1, x^2, x^2, x^3, 0 \right)
\end{equation}
(replacing the zero by a half does not lead to a fixed plane and
adding a one to any of the entries can be mapped to the above plane
with a different $x^3$ parameterisation by a lattice shift).

Finally, we consider the {\bf DDD} orientifold, i.e.\
\begin{equation}
{\cal R}: \,\,\, z^i \to \mbox{i}\overline{z}^i,\,\,\, i=1,2 ,3  .
\label{eq:AAa}
\end{equation}
Now, the ${\cal R}$ projected lattice is generated by
\begin{equation}
\Lambda_{{\cal R}, \perp}:\,\,\, \begin{array}{c} \left(
  \frac{1}{2},\frac{1}{2},\frac{1}{2},\frac{1}{2},0,0\right) ,\\
\left( 0,0,\frac{1}{2},\frac{1}{2},\frac{1}{2},\frac{1}{2}\right) , \\
\left( \frac{1}{2},\frac{1}{2},0,0,\frac{1}{2},\frac{1}{2}\right) ,
\end{array}
\end{equation}
where the first vector is obtained by acting with $\frac{{\cal R} +
  1}{2}$ on the SO(12) root $\left( 1,0,1,0,0,0\right)$. Any ${\cal
  R}$ invariant SO(12) root is on the $\Lambda_{{\cal R}, \perp}$
  lattice
  as it should be. The volume of $\Lambda_{{\cal R}, \perp}$ is
  $1/\sqrt{2}$.
The $-{\cal R}$ invariant lattice is spanned by
\begin{equation}
\Lambda_{-{\cal R}, inv}:\,\,\, \begin{array}{c} \left(
  1,-1,0,0,0,0\right) ,\\
\left( 0,0,1,-1,0,0\right) , \\
\left( 0,0,0,0,0,1,-1\right)
\end{array}
\end{equation}
and has volume $\sqrt{8}$. Computing volumes of very similar lattices we
obtain the tadpole cancellation for the {\bf DDD} orientifold
\begin{equation}
\left( 8 - N\right)^2 = 0.
\end{equation}
There are indeed two $\Omega {\cal R}$ fixed planes in this case
\begin{equation}
\left( x^1 ,x^1, x^2 ,x^2 ,x^3 ,x^3 \right) \,\,\, ,\,\,\, \left( x^1
+1, x^1, x^2 ,x^2, x^3 ,x^3\right) .
\end{equation}
Note that now the one in the second plane cannot be removed by a
lattice shift and a coordinate redefinition.

In summary, we find the tadpole cancellation conditions for the four
qualitatively different orientifolds of the SO(12) compactification:
\begin{equation}
  \begin{aligned}
    \text{{\bf CCC}:} & \qquad \left( N - 16\right)^2 = 0 \; ,\\
    \text{{\bf DCC}:} & \qquad \left( N - 8 \right)^2 = 0 \; ,\\
    \text{{\bf DDC}:} & \qquad \left( N - 4 \right)^2 = 0 \; ,\\
    \text{{\bf DDD}:} & \qquad \left( N - 8 \right)^2 = 0 \; .
  \end{aligned}
\label{eq:so12cases}
\end{equation}

\subsection{Adding D-branes at angles}\label{angles}

So far, we considered having only D6-branes parallel to
O6-planes. Open strings stretched between such branes yield non chiral
matter. In order to obtain a chiral
spectrum we have to add D6-branes forming non-trivial angles with the
O-planes.

For simplicity we focus here on branes which are invariant under the
orbifold action. (We comment on non invariant branes in
Sec.~\ref{nib}.) An orbifold invariant D-brane with label $a$ (or
stack of $N_a$ D-branes) wraps the three-cycle
\begin{equation} \label{eq:dgen}
D6_a = \left( m_a^1 \left[a_1\right] + n_a ^1 \left[ b_1\right]  \right)
\times \left( m_a ^2\left[ a_2\right] + n_a ^2 \left[
  b_2\right]\right) \times \left( m_a ^3\left[ a_3\right] + n_a ^3
\left[ b_3\right]\right) ,
\end{equation}
where we used the same notation for one-cycles as in the factorisable
literature (e.g.\ in \cite{Cvetic:2001tj}):
\begin{equation}
\begin{aligned}
\left[ a_1 \right] = \left( 1,0,0,0,0,0\right) ,\hspace*{1in} &
\left[ b_1 \right] = \left( 0,1,0,0,0,0\right) , \\
\left[ a_2 \right] = \left( 0,0,1,0,0,0\right) ,\hspace*{1in} &
\left[ b_2 \right] = \left( 0,0,0,1,0,0\right) , \\
\left[ a_3 \right] = \left( 0,0,0,0,1,0\right) ,\hspace*{1in} &
\left[ b_3 \right] = \left( 0,0,0,0,0,1\right) ,
\end{aligned}
\label{eq:halfyc}
\end{equation}
and $m^i _a$, $n_a ^i$ ($i=1,2,3$) are integers. The cycle
(\ref{eq:dgen}) is a closed cycle on the SO(12) compactification
lattice if
\begin{equation}\label{eq:so12cond}
m_a ^i + n_a ^i = \mbox{even}, \,\,\, i=1,2,3.
\end{equation}
In all other cases the D-brane has to wrap the cycle (\ref{eq:dgen})
twice in order to close on the SO(12) root lattice
compactification. This can be easiest seen for the case when
(\ref{eq:so12cond}) is violated only for one $i$. Then a closed
three-cycle is obtained by wrapping the corresponding one-cycle
twice. If instead (\ref{eq:so12cond}) is violated for $i=1,2$ we rewrite,
\begin{eqnarray}
2 \prod_{i=1}^3 \left( m^i_a \left[a_i\right] + n^i _a \left[
  b_i\right] \right) & = & \nonumber \\ & & \hspace*{-1.5in} \left( m_a
  ^1 , n_a ^1 , -m_a ^2, -n_a ^2, 0,0\right) \times \left( m_a ^1 ,
  n_a ^1 , m_a ^2 , n_a ^2 ,0,0\right) \times \left( 0,0,0,0, m_a ^3 ,
  n_a ^3\right) ,
\label{eq:twoodd}
\end{eqnarray}
where the r.h.s.\ clearly represents a closed three-cycle in the
SO(12) compactified case. (If both, $n_a ^3$ and $m_a ^3$, are even
then (\ref{eq:twoodd}) is actually twice a closed three-cycle.)
Eq.\ (\ref{eq:twoodd}) can be easily verified by employing a one-to-one
correspondence of homology and cohomology as
discussed, in the present context, e.g.\ in \cite{Villadoro:2006ia}.
Finally, if (\ref{eq:so12cond}) does not hold for any $i$ one notices
\begin{eqnarray}
2 \prod_{i=1}^3 \left( m^i_a \left[a_i\right] + n^i _a \left[
  b_i\right] \right) & = & \nonumber \\ & & \hspace*{-1.5in} \left(
  m_a ^1 , n_a ^1 , -m_a ^2, -n_a ^2,
  0,0\right) \times \left( m_a ^1 , n_a ^1 , m_a ^2 , n_a ^2
  ,0,0\right) \times \left( 0,0,m_a ^2,n_a ^2, m_a ^3 , n_a ^3\right) ,
\label{eq:threeodd}
\end{eqnarray}
implying that wrapping (\ref{eq:dgen}) twice suffices
to obtain a closed three-cycle on the SO(12) compactification
lattice\footnote{Similar arguments apply for other
  lattices. One can express (\ref{eq:threeodd}) in terms of
  fundamental three-cycles in the SO(12) root lattice. Then imposing
  co-prime conditions on the resulting expansion coefficients avoids
  multiple wrappings (counted already in $N_a$). }. 

In order to compute intersection numbers of two D6-branes one
determines a lattice in which the D-branes intersect once. The
Jacobian obtained when transforming to the compactification lattice
yields the intersection number \cite{Blumenhagen:2004di}.
To carry out the computation we express the three-cycle (\ref{eq:dgen})
in terms of SO(12) simple roots (\ref{eq:so12roots}):
\begin{eqnarray}
D6_a & = & \left( m^1_a e_1 +\left( m^1 _a + n^1_a\right) \left( e_2 + e_3
+ e_4 + \frac{1}{2} e_5 + \frac{1}{2} e_6 \right)\right)
\times
\nonumber \\ & & \hspace*{-0.5in}
\left( m_a ^2 e_3 +
\left( m_a ^2 + n_a ^2 \right) \left( e_4+ \frac{1}{2} e_5 +
\frac{1}{2} e_6\right)\right) \times \left( \frac{m_a ^3}{2} \left(
e_5 + e_6\right) + \frac{n_a ^3}{2} \left( e_6 - e_5\right)\right) .
\label{eq:dgeninsimp}
\end{eqnarray}
The lattice in which the brane $D6_a$ and a second brane $D6_b$
intersect once is spanned by the three one-cycles in
(\ref{eq:dgeninsimp}) and the corresponding one-cycles of $D6_b$. Thus
we obtain for their intersection number $I_{ab}$
\begin{eqnarray}
I_{ab} & = & \mbox{det} \left( \begin{array}{cccccc}
m_a ^1 & m_a ^1 + n_a ^1 & m_a ^1 + n_a ^1 & m_a ^1 + n_a ^1 & \frac{m_a
^1 + n_a ^1}{2} & \frac{m_a ^1 + n_a ^1}{2} \\
m_b ^1 & m_b ^1 + n_b ^1 & m_b ^1 + n_b ^1 & m_b ^1 + n_b ^1 & \frac{m_b
^1 + n_b ^1}{2} & \frac{m_b ^1 + n_b ^1}{2} \\
 0 & 0 & m_a ^2  & m_a ^2 + n_a ^2 & \frac{m_a
^2 + n_a ^2}{2} & \frac{m_a ^2 + n_a ^2}{2} \\
 0 & 0 & m_b ^2  & m_b ^2 + n_b ^2 & \frac{m_b
^2 + n_b ^2}{2} & \frac{m_b ^2 + n_b ^2}{2} \\
0 & 0 & 0 & 0 & \frac{m_a ^3 - n_a ^3}{2} & \frac{m_a ^3 + n_a ^3}{2}
\\
0 & 0 & 0 & 0 & \frac{m_b ^3 - n_b ^3}{2} & \frac{m_b ^3 + n_b ^3}{2}
\end{array}
\right) \nonumber \\
& & = \frac{1}{2} \prod_{i=1}^3 \left( m^i_a n_b ^i - n^i _a
m^i _b\right) .
\label{intso12}
\end{eqnarray}

In the previous section we identified the three-cycles wrapped by
O6-planes and checked that they are consistent with modular
transformations.
Similar to the D6-branes we can also express the three-cycles wrapped
by O6-planes in terms of the (half) cycles (\ref{eq:halfyc}) as,
\begin{equation}
\begin{array}{rl}
 O_{\Omega{\mathcal R}} &\hspace*{-0.1in} =  \left(
 1,0,\mbox{-1},0,0,0\right) \times \left(
0,0,1,0,\mbox{-1},0\right) \times
\left( 0,0,1,0,1,0\right) = 2 \left[a_1\right] \times \left[ a_2
  \right] \times \left[ a_3 \right],   \\
O_{\Omega{\mathcal R}\theta} &\hspace*{-0.1in} =  \left( 0,1 ,0 ,1
 ,0,0\right) \times
\left(  0,0,0,\mbox{-1,-1},0\right) \times
\left( 0,0,0,\mbox{-1},1,0\right)  = -2 \left[b_1\right] \times \left[ b_2
  \right] \times \left[ a_3 \right],  \\
O_{\Omega{\mathcal R}\omega} &\hspace*{-0.1in} =  \left(
 1,0,0,\mbox{-1},0,0\right) \times
\left(  0,0,0,1,0,1\right) \times
\left( 0,0,0,1,0,\mbox{-}1\right)  = -2 \left[a_1\right] \times \left[ b_2
  \right] \times \left[ b_3 \right], \\
O_{\Omega{\mathcal R}\theta\omega} &\hspace*{-0.1in} =  \left(
 0,1,\mbox{-1},0,0,0\right)
\times \left(  0,0,1,0,0,1\right) \times
\left( 0,0,1,0,0,\mbox{-1}\right)  = -2 \left[b_1\right] \times \left[ a_2
  \right] \times \left[ b_3 \right] .
\end{array}
\label{ocycles}
\end{equation}
Thus, as far as the RR tadpole cancellation is concerned, we view our
brane configuration as a compactification on $\left(T^2\right) ^3$
with the number of O-planes doubled (as compared to the SO(12) root
lattice). In the following we focus on the {\bf CCC} case discussed in
the previous section (see (\ref{eq:so12cases})). In this case there
are half as many O6-planes on the SO(12) lattice compactification as
in the {\bf AAA} $\left( T^2\right)^3$ compactification. According to
our discussion above we can, hence, just copy the tadpole cancellation
conditions from that example \cite{Blumenhagen:2000wh,Cvetic:2001tj}:
\begin{equation}
\begin{aligned}
\sum_a N_a m_a ^1 m_a ^2 m_a ^3 - 16 = & 0 , \\
\sum_a N_a m_a ^1 n_a ^2 n_a ^3 + 16 = & 0 , \\
\sum_a N_a n_a ^1 m_a ^2 n_a ^3 + 16 = & 0 , \\
\sum_a N_a n_a ^1 n_a ^2 m_a ^3 + 16 = & 0 .
\end{aligned}
\label{eq:tadpole}
\end{equation}

We look for solutions to (\ref{eq:tadpole}) which preserve
${\mathcal N}=1$ supersymmetry, i.e.\ the D-branes respect all the supersymmetry which
is unbroken by the orbifold and orientifold actions. That is, all
D-branes must be related to O-planes by SU(3) rotations commuting
with the orbifold group which is in the same SU(3)\cite{Berkooz:1996km}.
For concreteness, we consider different SU(2)
subgroups of SU(3) whose centers contain the orbifold elements  $\theta$,
$\omega$ or $\theta\omega$ (\ref{eq:orbaction}), respectively.
Explicitly, apart from
D-branes parallel to O-planes, we allow for  three types
of D6-branes to be present, if they wrap one of the following three-cycles
\begin{equation}
\begin{aligned}
(a) \qquad \left( k \left[ a_1\right] + l\left[ b_1\right] \right) \times \left(
m\left[ a_2\right] + n \left[ b_2 \right] \right) \times \left[ a_3
  \right] ,& \\
(b) \qquad\left[ a_1
  \right]\times \left( k \left[ a_2\right] + l\left[ b_2\right]
\right) \times \left(
m\left[ a_3\right] + n \left[ b_3 \right] \right)  ,& \\
(c) \qquad\left( k \left[ a_1\right] + l\left[ b_1\right] \right) \times\left[ a_2
  \right] \times  \left(
m\left[ a_3\right] + n \left[ b_3 \right] \right) , &
\end{aligned}
\end{equation}
with
\begin{equation}
\frac{m}{n} = - \frac{k}{l}.
\end{equation}
For consistency one has to add also the orientifold images, i.e.\
branes for which the signs of $n$ and $l$ are reversed.

Although the tadpole cancellation conditions are rather restrictive
one can find a chiral model with several non abelian gauge factors.
We list the corresponding D-branes in Table
\ref{tab:chiralbranes}.
\begin{table}[h!]
\begin{center}
\begin{tabular}{|c|| c | r | r | r | r | r | r |}
\hline \hline
Type &  $N_a$ & $m_a ^1 $ & $n_a ^1$ & $m_a ^2$ & $n_a ^2$ & $m_a ^3$
& $n_a ^3$ \\ \hline
$A_1$ & 6 & 1 & 1 & 1 & $-1$ & 2 & 0 \\
$B_1$ & 2  & 2 & 0 & 1 & 1  & 1 & $-1$ \\
\hline
$P_1$ & 2 & 0 & 1 & 0 & $-1$ & 2 & 0 \\
$ P_2 $ & 6 & 2 & 0 & 0 & 1 & 0 & $-1$ \\
$P_3$ & 8 & 0 & 1 & 2 & 0 & 0 & $-1$ \\
\hline \hline
\end{tabular}
\end{center}
\caption{\it D6-brane configuration for a chiral supersymmetric model. As
  explained
  in the text, the choice of wrapping numbers $m_a ^i$, $n_a ^i$
  yields one closed three-cycle on the SO(12) lattice
  compactification, for each type of D6-branes. The number $N_a$
  denotes the number of D6-branes in a stack. \label{tab:chiralbranes}}
\end{table}
The chiral part of the spectrum is given in Table \ref{tab:chispec}
(the detailed rules for computing this spectrum from given
intersection numbers can be found in \cite{Cvetic:2001tj}, see also
Appendix \ref{ap:rules}). 
\begin{table}[h!]
\begin{center}
\begin{tabular}{| c ||c |r |}
\hline \hline
Sector & U(3)$\times$USp(2)$\times$USp(6)$\times$USp(8) & Q\\ \hline
$A_1 B_1$ & $ 4\left( 3,1,1,1\right)$ &$ -1$ \\
$A_1\left( P_2 + P_2^\prime\right) $&  $2 \left( 3,1,6,1\right) $ & 0 \\

$A_1 \left( P_3 + P_3 ^\prime\right)$ &  $2 \left(
\overline{3},1,1,8\right)$ & 0 \\
$B_1 \left( P_1 + P_1 ^\prime\right)$ & $ 2\left( 1,2,1,1\right)$ & $-1$
\\
$B_1 \left( P_3 + P_3 ^\prime\right)$ &  $2\left( 1,1,1,8\right) $ & 1
\\
\hline \hline
\end{tabular}
\end{center}
\caption{\it Chiral spectrum of the configuration in Table
  \ref{tab:chiralbranes}. Here, we adopted the convention that Eq.\
  (B2) of \cite{Cvetic:2001tj} yields chiral multiplets in the given
  representations. $Q$ denotes the charge under the U(1) living on the
  stack $B_1$. The rules are summarised in Appendix \ref{ap:rules}.
  \label{tab:chispec}} 
\end{table}

The brane configuration in Table \ref{tab:chiralbranes} is suitable to
visualise the effect of non factorisable as compared to factorisable
compactifications. One can start with the same brane configuration in
ten dimensions but instead of compactifying on the non-factorisable
SO(12) lattice take the factorisable {\bf AAA} lattice. Now, the
branes in Table \ref{tab:chiralbranes} wrap closed three-cycles twice
and one obtains a modified Table \ref{tab:factorbr}.
\begin{table}[h!]
\begin{center}
\begin{tabular}{|c|| c | r | r | r | r | r | r |}
\hline \hline
Type &  $N_a$ & $m_a ^1 $ & $n_a ^1$ & $m_a ^2$ & $n_a ^2$ & $m_a ^3$
& $n_a ^3$ \\ \hline
$A_1$ & 12 & 1 & 1 & 1 & $-1$ & 1 & 0 \\
$B_1$ & 4  & 1 & 0 & 1 & 1  & 1 & $-1$ \\
\hline
$P_1$ & 4 & 0 & 1 & 0 & $-1$ & 1 & 0 \\
$ P_2 $ & 12 & 1 & 0 & 0 & 1 & 0 & $-1$ \\
$P_3$ & 16 & 0 & 1 & 1 & 0 & 0 & $-1$ \\
\hline \hline
\end{tabular}
\end{center}
\caption{\it D6-brane configuration for a chiral supersymmetric model in a
  factorisable {\bf AAA} setting. Now, the choice of wrapping numbers
  $n_a ^i$, $m_a ^i$
  yields one closed three-cycle on the {\bf AAA} lattice
  compactification, for each type of D6-branes. In ten dimensions the
  branes extend along the same directions as the ones in Table
  \ref{tab:chiralbranes}.\label{tab:factorbr} }
\end{table}
The corresponding chiral part of the massless spectrum is listed in
Table \ref{tab:specfac}.
\begin{table}[h!]
\begin{center}
\begin{tabular}{| c ||c |r |}
\hline \hline
Sector & U(6)$\times$U(2)$\times$USp(4)$\times$USp(12)$\times$USp(16) \\
\hline  & \\[-2ex]
$A_1 B_1$ & $ 2\left( 6,\overline{2},1,1,1\right)$  \\
$A_1\left( P_2 + P_2^\prime\right) $&  $1 \left( 6,1,1,12,1\right) $  \\
$A_1 \left( P_3 + P_3 ^\prime\right)$ &  $1 \left(
\overline{6},1,1,1,16\right)$  \\
$B_1 \left( P_1 + P_1 ^\prime\right)$ & $ 1\left(
1,\overline{2},4,1,1\right)$
\\
$B_1 \left( P_3 + P_3 ^\prime\right)$ &  $1\left( 1,2,1,1,16\right) $
\\
\hline \hline
\end{tabular}
\end{center}
\caption{\it Chiral spectrum of the configuration in Table
  \ref{tab:factorbr}. Comparison with Table
  \ref{tab:chispec} shows that in the factorisable case the rank of
  the gauge group factors is twice as big whereas the number of
  generations is half the numbers obtained for the SO(12)
  compactification.\label{tab:specfac}}
\end{table}
We see that replacing a factorisable compactification by a non
factorisable one decreases the size of the gauge group whereas it
increases the number of generations. Obviously the two models cannot
be connected by conventional continuous deformations like spontaneous
symmetry breaking by turning on flat directions in the moduli space.


\section{Number of families}

\subsection{Factorisable lattices}

In this section we review the simple argument of why  one cannot
obtain an odd number of generations on the factorisable {\bf AAA}
torus and how introducing a tilted torus solves the problem. On the
{\bf AAA} torus the branes wrap the cycles (\ref{eq:dgen})

\begin{equation}
D6_a = \left( m_a^1 \left[a_1\right] + n_a ^1 \left[ b_1\right]  \right)
\times \left( m_a ^2\left[ a_2\right] + n_a ^2 \left[
  b_2\right]\right) \times \left( m_a ^3\left[ a_3\right] + n_a ^3
\left[ b_3\right]\right) , \nonumber
\end{equation}
where the one-cycles are defined in (\ref{eq:halfyc}), while the image
cycles under $\Omega {\cal R}$ are obtained by replacing $n_a^i$ with
$-n_a^i$. Hence the intersection numbers are given by
\bea
I_{ab}&=&\prod_{i=1}^3 ( m^i_a n_b ^i - n^i _a
m^i _b) ,\nonumber\\
I_{ab'}&=&-\prod_{i=1}^3 ( m^i_a n_b ^i + n^i _a
m^i _b) .
\label{images}
\eea

If the stack $D6_a$ generates the U(3) gauge group and the stack
$D6_b$ gives the U(2) gauge group, in order to have three copies of
the $(3,2)$ representation of SU(3)$\times$ SU(2) we need either (i)
$I_{ab}=3$ and $I_{ab'}=0$ or (ii) $I_{ab}=2$ and
$I_{ab'}=1$. One could also have only the net number of
generations to be three, e.g.\ four generations and one
anti-generation. However, from  (\ref{images}), it follows that
\begin{eqnarray}
I_{ab}+I_{ab'} & = & \nonumber\\ & &
\hspace*{-1in}-2\left[m_a^1m_a^2n_a^3 n_b^1n_b^2m_b^3 +
  m_a^1n_a^2m_a^3n_b^1m_b^2n_b^3 +
        n_a^1n_a^2n_a^3m_b^1m_b^2m_b^3
+ n_a^1m_a^2m_a^3m_b^1n_b^2n_b^3\right]
\label{family}
\end{eqnarray}
is even for any values of the integer wrapping numbers\cite{Blumenhagen:2000wh}. The solution proposed in \cite{Blumenhagen:2000ea} to solve this problem, consists in tilting one of the three two dimensional tori. This amounts to replacing the {\bf A} lattice in one of the tori with a {\bf B} lattice  (see Sec.~\ref{recap})
\bea
e_1&=&(1, -1, 0, 0, 0,0) ,\nonumber\\
e_2&=&(1,~1,0,0, 0,0) ,\nonumber\\
e_3&=&(0, 0,1 , 0, 0,0) ,\nonumber\\
e_4&=&(0, 0,0 , 1, 0,0) ,\nonumber\\
e_5&=&(0, 0,0 , 0, 1,0) ,\nonumber\\
e_6&=&(0, 0,0 , 0, 0,1) .
\label{baa}
\eea
Expressing the three-cycle (\ref{eq:dgen}) in terms of the new lattice basis vectors we obtain
\begin{equation}
D6_a = \left( \frac{m_a^1}{2}~(e_1+e_2) + \frac{n_a ^1}{2} ~(e_2-e_1) \right)
\times \left( m_a ^2 ~e_3 + n_a ^2~ e_4 \right) \times \left( m_a ^3~e_5 + n_a ^3
~e_6\right) ,
\label{dtt}
\end{equation}
which describes a closed cycle on the {\bf BAA} lattice (\ref{baa}) if

\be m_a^1+n_a^1={\rm even}.\label{condbaa}\ee
 Equally the intersection number is modified to
\be
I_{ab}=\frac{1}{2}\prod_{i=1}^3 ( m^i_a n_b ^i - n^i _a
m^i _b)\label{intbaa}
\ee
and therefore, the factor of $2$ in front of (\ref{family}) cancels off.
Alternatively, the condition (\ref{condbaa}) introduces a minimal
factor of 2 if $m_a^1$ and $n_a^1$ are both odd (and a minimal factor
of 4 if $m_a^1,~n_a^1$ are even), which is compensated by the global
factor in the intersection number (\ref{intbaa}).
Thus, in this case three generations can be obtained in both ways
 described at the beginning of the section \cite{Cvetic:2001tj,cll}
(see Table \ref{ex3genfact} for some examples).
\begin{table}[h!]
\begin{center}
\begin{tabular}{|c||c|c | r | r | r | r | r | r |}
\hline \hline
Case & Type &  $N_a$ & $m_a ^1 $ & $n_a ^1$ & $m_a ^2$ & $n_a ^2$ & $m_a ^3$
& $n_a ^3$ \\ \hline
(i) & $A$ & 6 & 3 & 1 & 1 & $-1$ & 1 & 0 \\
~& $B$ & 4  & 1 & 1 & 1 & 0  & 1 & $-1$ \\
\hline\hline
(ii) &$A$ & 6 & 1 & 1 & 0 & $-1$ & 1 & 1 \\
~& $B$ & 4  & 1 & $-1$ & 3 & 1  & 1 & 0 \\
\hline \hline
\end{tabular}
\end{center}
\caption{\it Examples with three generations on factorisable lattices.}
\label{ex3genfact}
\end{table}

\subsection{Non factorisable lattices}

Let us now consider a lattice similar to (\ref{baa}), but which does
not factorise under the action of the ${\mathbb Z}_2 \times {\mathbb
  Z}_2$ orbifold
\bea
e_1&=&(1, 0, -1, 0, 0, 0) ,\nonumber\\
e_2&=&(0,1,0,0, 0,0) ,\nonumber\\
e_3&=&(1, 0,1 , 0, 0,0) ,\nonumber\\
e_4&=&(0, 0,0 , 1, 0,0),\nonumber\\
e_5&=&(0, 0,0 , 0, 1,0),\nonumber\\
e_6&=&(0, 0,0 , 0, 0,1).
\label{baanfact}
\eea
The three-cycle (\ref{eq:dgen}) now takes the form
\begin{equation}
D6_a = \left( \frac{m_a^1}{2}~(e_1+e_3) +n_a ^1~ e_2 \right)
\times \left( \frac{m_a ^2}{2}~( e_3-e_1) + n_a ^2 ~e_4 \right) \times \left( m_a ^3 ~e_5 + n_a ^3~
e_6\right) ,
\end{equation}
The intersection number is given again by (\ref{intbaa}), but
 the condition (\ref{condbaa}) becomes $\footnote{Note that in order to keep (\ref{condbaa}) unchanged one would need to consider branes of the form $(m_a^1,0,n_a^1,0,0,0) \times (0,m_a^2,0,n_a^2,0,0) \times (0,0,0,0,m_a^3,n_a^3)$, which are not orbifold invariant. This possibility is discussed in the section \ref{nib}.}$
\be\label{condinf}
m_a^1\,, ~m_a^2={\rm even}.
\ee
If these conditions are satisfied, each introduces a minimal factor of
two in the intersection number, hence there is a factor of two too many
and we cannot obtain odd intersection numbers in this case.
If one or both  conditions above are violated, however, the situation
is similar to the case we studied in  Sec.~\ref{angles}. That is, the
brane has to wrap twice the cycle (\ref{eq:dgen}). In that case, it
is possible to get odd intersection numbers
$I_{ab}$. This can be achieved by intersecting branes of two
different types. For one type the wrapping numbers $m^1$,
$m^2$ are both odd (and one has to wrap the cycle twice), and for the
other type both are even. However the total number of families, given by
 (\ref{family}), is always even for any combination of branes obeying,
 or not,  the conditions
(\ref{condinf}).

 In the following we argue that this happens for a general choice of
 non factorisable lattices. In the example at hand one can see that
 the condition (\ref{condbaa}) is modified because, in the non
 factorisable case, the coordinates $x^1$ and $y^1$ are not related anymore in
 the basis vectors. Nevertheless the even intersection number problem
 remains also when the condition (\ref{condbaa}) is preserved.

If we take the example of the SO(12) root lattice, the condition for
having  closed cycles of the form (\ref{eq:dgeninsimp}) is $m_a ^i +
n_a ^i = \mbox{even}, \,\,\, i=1,2,3$, similar to the factorisable
case. If this condition is satisfied for all $i$'s, each  contributes
with a minimal factor of 2. Thus the single 1/2 factor in the
intersection number cannot account for them. As we saw in
Sec.~\ref{angles}, if this condition is not satisfied for one or all
$i$'s, the brane has to wrap the cycle twice. It is again possible to
check that, although the single intersection numbers can
be odd, the total intersection number
$I_{ab} + I_{ab'}$  is always even.

In contrast, in the factorisable case with three tilted tori, i.e.~an
SO(4)$^3$ factorised lattice,
\begin{equation}
\begin{aligned}
 e_1 &=& \left( 1,-1,0,0,0,0\right) ,\\
 e_2  &=& \left( 1,~~1,0,0,0,0\right) , \\
 e_3 &=& \left( 0,0,1,-1,0,0\right) ,\\
 e_4 &=& \left( 0,0,1,~~1,0,0\right) , \\
 e_5  &=& \left( 0,0,0,0,1,-1\right) ,\\
 e_6  &=& \left( 0,0,0,0,1,~~1\right) ,
\end{aligned}
\label{so4fact}
\end{equation}
although we also have three conditions of the form $m_a ^i + n_a ^i =
\mbox{even}, \,\,\, i=1,2,3$, the intersection number also contains,
this time, three factors of 1/2. This can be seen from the
intersection number. While in the SO(12) case, equation
(\ref{intso12}), there is only a single 1/2 factor that survives after
performing operations that leave the determinant invariant, like
adding columns, in the case of the factorised SO(4)$^3$ lattice we
have
\begin{eqnarray}
I_{ab} & = & \mbox{det} \left( \begin{array}{cccccc}
\frac{m_a ^1-n_a^1}{2} & \frac{m_a ^1+n_a^1}{2} & 0 & 0 & 0 & 0 \\
\frac{m_b^1 - n_b ^1}{2} & \frac{m_b^1 + n_b ^1}{2} & 0 & 0 & 0 & 0\\
 0 & 0 & \frac{m_a ^2 - n_a ^2}{2}  & \frac{m_a ^2 + n_a ^2}{2} & 0 & 0 \\
 0 & 0 & \frac{m_b ^2 - n_b ^2}{2}  & \frac{m_b ^2 + n_b ^2}{2} & 0 & 0 \\
0 & 0 & 0 & 0 & \frac{m_a ^3 - n_a ^3}{2} & \frac{m_a ^3 + n_a ^3}{2}
\\
0 & 0 & 0 & 0 & \frac{m_b ^3 - n_b ^3}{2} & \frac{m_b ^3 + n_b ^3}{2}
\end{array}
\right) \nonumber \\
& & = \frac{1}{8} \prod_{i=1}^3 \left( m^i_a n_b ^i - n^i _a
m^i _b\right) .
\end{eqnarray}
And therefore, it is possible to get an odd number of
families\footnote{Although, as pointed out in \cite{cll}, other
  phenomenological requirements eliminate the possibility to get
  consistent models using more than one tilted torus. }.

In the non factorisable case, as well, one can obtain a factor of 1/8 in the intersection number. Using again an  SO(4)$^3$ lattice as an example, we can take, for instance
\begin{equation}
\begin{aligned}
 e_1 &=& \left( 1,0,-1,0,0,0\right) ,\\
 e_2  &=& \left( 1,0,~~1,0,0,0\right) , \\
 e_3 &=& \left( 0,1,0,0,-1,0\right) ,\\
 e_4 &=& \left( 0,1,0,0,~~1,0\right) , \\
 e_5  &=& \left( 0,0,0,1,0,-1\right) ,\\
 e_6  &=& \left( 0,0,0,1,0,~~1\right) .
\end{aligned}
\label{so4nonfact}
\end{equation}
In this case, the invariant cycles (\ref{eq:dgen}) take the form
\begin{eqnarray}
D6_a & = & \left( \frac{m^1_a}{2} \left( e_1 +e_2 \right)+\frac{n_a^1}{2}\left(e_3 +e_4\right)\right)
\times
\left( \frac{m_a ^2}{2} \left(e_2 -e_1\right) + \frac{n_a ^2}{2} \left(e_5 +e_6\right)\right)  \times  \nonumber \\
& &\left( \frac{m_a ^3}{2} \left(
e_4 - e_3\right) + \frac{n_a ^3}{2} \left( e_6 - e_5\right)\right),
\label{eq:so4nonfact}
\end{eqnarray}
which translates into the intersection number
\begin{eqnarray}
I_{ab} & = & \mbox{det} \left( \begin{array}{cccccc}
\frac{m_a ^1}{2} & \frac{m_a ^1}{2} & \frac{n_a^1}{2} & \frac{n_a^1}{2} & 0 & 0 \\
\frac{m_b^1}{2} & \frac{m_b^1 }{2} & \frac{n_b ^1}{2}  & \frac{n_b ^1}{2}  & 0 & 0\\
-\frac{m_a ^2 }{2} & \frac{m_a ^2}{2} & 0  & 0 & \frac{n_a ^2}{2} & \frac{n_a ^2}{2} \\
-\frac{m_b ^2 }{2} & \frac{m_b ^2}{2} & 0  & 0 & \frac{n_b ^2}{2} & \frac{n_b ^2}{2} \\
0 & 0 & -\frac{m_a ^3 }{2} & \frac{m_a ^3 }{2} & -\frac{n_a ^3}{2} & \frac{ n_a ^3}{2}
\\
0 & 0 & -\frac{m_b ^3 }{2} & \frac{m_b ^3 }{2} & -\frac{n_b ^3}{2} & \frac{ n_b ^3}{2}
\end{array}
\right) \nonumber \\
& & = \frac{1}{8} \prod_{i=1}^3 \left( m^i_a n_b ^i - n^i _a
m^i _b\right) ,
\end{eqnarray}
but with the conditions $m_a ^i= \mbox{even}, ~n_a ^i = \mbox{even}, \,\,\, i=1,2,3$. Taking into account the possibility of taking non-closed cycles (following the rules discussed in Sec.~\ref{angles}) in total, these conditions would introduce a minimal factor of $4$ too much. It seems difficult to reduce the number of conditions to one, while maintaining a factor of 1/8, or at least 1/4, in the intersection  number.

 From the point of view of minimising the factors of 2, conditions of the form (\ref{condbaa}) seem to be preferable. But these conditions seem to be correlated with less global factors of one half (a single factor of 1/2 in the examples above). So, one should try to reduce the  number of conditions to one, in a non factorisable way, for instance
\begin{equation}\label{onempncond}
\begin{aligned}
e_1 = & \left( 1,-1,0,0,0,0\right) ,\\
e_2 = & \left( 0, 1 , -1,0,0,0\right) , \\
e_3 = & \left( 0,1,~1,0,0,0\right) , \\
e_4 = & \left( 0,0,0,1 ,0,0\right) \\
e_5 = & \left( 0,0,0,0,1,0\right) ,\\
e_6 = & \left( 0,0,0,0,0, 1\right) ,
\end{aligned}
\end{equation}
where
\begin{eqnarray}
D6_a & = & \left( m^1_a \left( e_1 +\frac{e_2 +e_3}{2}\right)+\frac{n_a^1}{2}\left(e_2 +e_3\right)\right)
\times \nonumber \\
& &\left( \frac{m_a ^2}{2} \left(e_3 -e_2\right) + n_a ^2e_4 \right)  \times
\left( m_a ^3e_5 + n_a ^3 e_6 \right).
\label{eq:onempn}
\end{eqnarray}
The price of realising condition (\ref{condbaa}) in a non
factorisable way is to have also $m_a^2=\mbox{even}$, while there is
still just a single factor of one half in the intersection number. One
can think of other examples, but each time the conditions for having
closed cycles introduce at least a factor of two too many.

The only other choice one can think of is to consider a different
orientifold action, like the one in equation (\ref{eq:ABB})
\begin{equation}
{\cal R}: \,\,\, z^1 \to \mbox{i}\overline{z}^1,\,\,\, z^i \to
\overline{z}^i, \,\,\, i=2,3 .\nonumber
\end{equation}
The advantage of this action is that the image branes are not obtained
by replacing $n$ with $-n$, but $m$ with $n$ and vice versa, which
avoids having even $I_{ab}+I_{ab'}$ in all  cases. On the other
hand the number of lattices that admit this symmetry is
reduced. Particularly interesting in this case are the conditions that
restrict only one of the wrapping numbers, say
$m_a^i=\mbox{even},~~i=1,2,3$, since
\be
I_{ab'}\sim\prod_{i=1}^3 \left( m^i_a m_b ^i - n^i _a
n^i _b\right).
\ee
Actually in order to have conditions only on the wrapping numbers $m$
we need to have a factorised lattice in the coordinates $y^1$, $y^2$
and $y^3$, but
the symmetry (\ref{eq:ABB}) relates the coordinates $x^1$ and $y^1$. So, the
condition $m_a^1=\mbox{even}$ and the (\ref{eq:ABB})  are not
compatible.
The other type of conditions, $m_a^i,~n_a^i=\mbox{even}$ and $m_a^i+n_a^i=\mbox{even}$,  do not make a difference with the previous case ($I_{ab}+I_{ab'}$ is again even).

\subsection{Non-invariant branes}\label{nib}

To complete our search for models in non-factorisable tori,  in this
section we study the possibility of adding non-invariant branes
under the orbifold group. 
For instance, consider a pair of branes wrapping the following cycles 
\begin{equation}
\begin{aligned}
D6_a & = & (1, 0, 0, 0, 0, 0)\times(0, 0, 0, 1, 3, 0)\times(0, 0, 1, 0,
0, -1) ,\\
D6_b & = & (1, -1, 0, 0, 0, 0)\times(0, 0, 0, 1, 1, 0)\times(0, 0, 1, 0,
0, 0) .
\end{aligned}
\label{rotated}
\end{equation}
These branes are not invariant under the orbifold action.  Moreover,
they are rotated with respect to the O6-planes, but along non standard
directions.  For example, if the six dimensional torus has complex
coordinates  $z^i = x^i+$i$ y^i$, $i=1,\,2,\,3$, the first brane
above can be put in an invariant form by rotating it by $\pm \pi/2$  in
the plane $(x^2,\,x^3)$. Thus such brane would be  related to the
O6-planes by rotations which do not commute with the orbifold.
Therefore, branes of type (\ref{rotated}),  do not preserve any
supersymmetry. In spite of this, one can check that this kind of
configurations give rise to  an odd number of families. More
specifically, from the configuration above one gets
 $I_{ab} + I_{ab'} =3$ \footnote{This can be realised on a lattice
  similar to (\ref{baanfact}), but with the non factorisable lattice
  in the coordinates $y^2$ and $x^3$, instead of $x^1$ and
  $x^2$. Intersection points which are not invariant under the
  orbifold are mapped onto intersection points of image branes.}.
Thus, even if these configurations do not preserve supersymmetry,
one could still construct chiral four dimensional  models, which
might give spectra close to that of the Standard Model. One would
have to check if there is a way to make such models stable or long lived.

\section{Conclusions}

In this note, we have explored in detail orientifold models of Type
IIA string theory compactified on non-factorisable lattices.
We have concentrated in orientifolds of $T^6/{\mathbb Z_2}\times
{\mathbb Z_2}$, which admit more general lattices. In particular, we
were interested in lattices that cannot be expressed in a factorisable
fashion.
Initial work along these lines, was started in
\cite{Blumenhagen:2004di}. There the authors concentrated in
orientifolds of
$T^6/{\mathbb Z_N}$, and restricted their study to
non-chiral four dimensional models.

We have taken a step further and considered the possibility of
including D6-branes at angles, which can then give rise to chiral
models in four dimensions.
We did this by working explicitly with an illustrative
example, the SO(12) lattice. As we saw, once one introduces
non-factorisable lattices, the tadpoles conditions change according to
the lattice. As expected, lattice vectors forming non trivial angles
with Euclidean coordinate axes lead to rank reductions in the gauge
symmetries.
Moreover, we saw that consistency with the compactification imposes
strong constraints on the wrapping numbers of the D6-branes, $(m,n)$.
These conditions  get reflected in the intersection numbers, which are
directly connected to the number of families.
As we showed, when one considers orbifold invariant branes,
the total number of families, which is given by $I_{ab} + I_{ab'}$
turns out to be always even, whether supersymmetry is imposed or not.

In the case of non-invariant branes,  it is supersymmetry which
forbids to get an odd  number of families.
However, as we saw in the last section, non-supersymmetric models with
odd number of families can be constructed,
although their stability might be a problematic issue.

Thus, our findings seem to imply a  dramatic conclusion. The models in
\cite{Cvetic:2001tj} appear to be, as the authors stated,  quite
unique. Unfortunately, all those models suffer from the presence of
several chiral exotic particles in their spectra.
Therefore, one would be tempted to conclude that supersymmetric
orientifold models on $T^6/{\mathbb Z_2}\times \mathbb Z_2$ are not
viable phenomenologically.
A possibly related observation has been reported in
\cite{schel-pheno}. In an approach along the lines of
\cite{Dijkstra:2004ym}, with the CFT given by free fermions, the
collaboration could exclude all models phenomenologically. It is not
clear that there should be a connection to our results. In the context
of heterotic constructions, it has been conjectured, however, that
semi realistic free fermionic models and orbifolds of non factorisable
six-tori are related \cite{Faraggi:1993pr}. Therefore, there might be
some correlation between our results and those of \cite{schel-pheno}.

It would be interesting to explore chiral model constructions in
other orientifolds, for instance those discussed in
\cite{Blumenhagen:2004di}, to see if the situation can be
ameliorated. A possibility to improve the situation for ${\mathbb Z}_2
\times {\mathbb Z}_2$ may be to include projections acting
geometrically as free shifts as described in \cite{Donagi:2004ht} for
heterotic orbifolds and in \cite{Blumenhagen:2006ab} for orientifolds.

Another issue that we did not touch at this level of our discussion,
is the problem of moduli stabilisation. In the context of factorisable
tori, this issue has been investigated in \cite{ms}. However, as in
the case of \cite{Cvetic:2001tj}, such models have chiral exotic
fields.

\section*{Acknowledgments}

We thank Ralph Blumenhagen and Gabriele Honecker for discussions at
the String Pheno 07 conference where parts of our results have been
presented. SF acknowledges the kind hospitality extended towards him
during visits at Liverpool University.
IZ is supported by a PPARC Postdoctoral Fellowship.

\begin{appendix}

\section{Model building rules \label{ap:rules}}

Here, we summarise the model building rules for the $T^6/\left(
{\mathbb Z}_2 \times {\mathbb Z}_2\right)$ orientifold. A stack of $N$
D6-branes not situated on top of an O6-plane accommodates the gauge
symmetry $U\left( N/2\right)$. If $N$ D6-branes are located on top of an
O6-plane they give 
rise to the gauge factor $USp\left(N\right)$. The chiral spectrum comes from
strings stretched between branes tilted with respect to the
O-planes. The rules are \cite{Blumenhagen:2000wh} ($a \not= b$):
\begin{itemize}
\item Strings stretching between the brane-stacks $N_a$ and $N_b$
  give rise to
  $I_{ab}$ multiplets in the $\left( \frac{N_a}{2},
  \frac{\overline{N}_b}{2}\right)$ 
  representation of $U\left(N_a/2\right) \times U\left( N_b/2\right)$.
\item 
 Strings stretching between the brane-stack $N_a$ and the ${\cal
  R}$-image-stack $N_{b^\prime}$ yield
  $I_{ab^\prime }$ multiplets in the $\left( \frac{N_a}{2},
  \frac{N_{b^\prime}}{2}\right)$ 
  representation of $U\left(N_a/2\right) \times U\left( N_{b^\prime}/2\right)$.
\item
Strings stretching between the brane-stack $N_a$ and its ${\cal R}$
  image provide 
  $\frac{1}{2}\left( I_{aa^\prime } + 4 I_{aO6}\right)$ multiplets in
  the anti-symmetric 
  representation of $U\left(N_a/2\right)$, and
  $\frac{1}{2}\left(I_{aa^\prime } - 4 I_{aO6}\right)$ in the symmetric
  representation. Here, O6 refers to the sum of all three-cycles
  wrapped by O6-planes. 
\end{itemize} 
In addition there is non-chiral matter.
\end{appendix}

\end{document}

%% file: 2dfix.pstex_t
\begin{picture}(0,0)%
\includegraphics{2dfix.pstex}%
\end{picture}%
\setlength{\unitlength}{1302sp}%
\begingroup\makeatletter\ifx\SetFigFont\undefined%
\gdef\SetFigFont#1#2#3#4#5{%
  \reset@font\fontsize{#1}{#2pt}%
  \fontfamily{#3}\fontseries{#4}\fontshape{#5}%
  \selectfont}%
\fi\endgroup%
\begin{picture}(12878,8691)(388,-8194)
\put(9016,-181){\makebox(0,0)[lb]{\smash{{\SetFigFont{11}{13.2}{\rmdefault}{\mddefault}{\updefault}{\bf B} lattice}}}}
\put(1501,-136){\makebox(0,0)[lb]{\smash{{\SetFigFont{11}{13.2}{\rmdefault}{\mddefault}{\updefault}{\bf A} lattice}}}}
\end{picture}%

%% file: FTZv3.bbl
\begin{thebibliography}{99}
%
\bibitem{Sagnotti:1987tw}
  A.~Sagnotti,
  arXiv:hep-th/0208020;
  G.~Pradisi and A.~Sagnotti,
  Phys.\ Lett.\  B {\bf 216} (1989) 59;
  M.~Bianchi and A.~Sagnotti,
  Phys.\ Lett.\  B {\bf 247} (1990) 517.
%
\bibitem{Horava:1989vt}
  P.~Ho\v{r}ava,
  Nucl.\ Phys.\  B {\bf 327} (1989) 461.
%
\bibitem{Govaerts:1988md}
  J.~Govaerts,
  Phys.\ Lett.\  B {\bf 220} (1989) 77.
%
\bibitem{Bern:1989zr}
  Z.~Bern and D.~C.~Dunbar,
  Phys.\ Rev.\ Lett.\  {\bf 64} (1990) 827.
%
\bibitem{Gimon:1996rq}
  E.~G.~Gimon and J.~Polchinski,
  Phys.\ Rev.\  D {\bf 54} (1996) 1667
  [arXiv:hep-th/9601038].
%
\bibitem{Blumenhagen:2000wh}
  R.~Blumenhagen, L.~G\"orlich, B.~K\"ors and D.~L\"ust,
  JHEP {\bf 0010} (2000) 006
  [arXiv:hep-th/0007024].
%
\bibitem{imr}
  L.~E.~Ib\'a\~{n}ez, F.~Marchesano and R.~Rabad\'an,
  JHEP {\bf 0111} (2001) 002
  [arXiv:hep-th/0105155].
%
\bibitem{Blumenhagen:2001te}
  R.~Blumenhagen, B.~K\"ors, D.~L\"ust and T.~Ott,
  Nucl.\ Phys.\  B {\bf 616} (2001) 3
  [arXiv:hep-th/0107138];
  R.~Blumenhagen, B.~K\"ors and D.~L\"ust,
  Phys.\ Lett.\  B {\bf 532} (2002) 141
  [arXiv:hep-th/0202024].
%
\bibitem{Angelantonj:2000hi}
  C.~Angelantonj, I.~Antoniadis, E.~Dudas and A.~Sagnotti,
  Phys.\ Lett.\  B {\bf 489}, 223 (2000)
  [arXiv:hep-th/0007090];
  C.~Angelantonj, M.~Cardella and N.~Irges,
  Nucl.\ Phys.\  B {\bf 725} (2005) 115
  [arXiv:hep-th/0503179].
%
\bibitem{Aldazabal:2000cn}
  G.~Aldazabal, S.~Franco, L.~E.~Ib\'{a}\~{n}ez, R.~Rabad\'an and
  A.~M.~Uranga,
  JHEP {\bf 0102} (2001) 047
  [arXiv:hep-ph/0011132];
  J.\ Math.\ Phys.\  {\bf 42} (2001) 3103
  [arXiv:hep-th/0011073].
%
\bibitem{Forste:2001gb}
  S.~F\"orste, G.~Honecker and R.~Schreyer,
  JHEP {\bf 0106} (2001) 004
  [arXiv:hep-th/0105208];
  G.~Honecker,
  JHEP {\bf 0201} (2002) 025
  [arXiv:hep-th/0201037].
%
\bibitem{Bailin:2001ie}
  D.~Bailin, G.~V.~Kraniotis and A.~Love,
  Phys.\ Lett.\  B {\bf 530} (2002) 202
  [arXiv:hep-th/0108131];
  Phys.\ Lett.\  B {\bf 547} (2002) 43
  [arXiv:hep-th/0208103];
  Phys.\ Lett.\  B {\bf 553} (2003) 79
  [arXiv:hep-th/0210219];
%
  JHEP {\bf 0302} (2003) 052
  [arXiv:hep-th/0212112].
%
  \bibitem{koko}
  C.~Kokorelis,
  JHEP {\bf 0208} (2002) 018
  [arXiv:hep-th/0203187];
  JHEP {\bf 0209} (2002) 029
  [arXiv:hep-th/0205147];
  JHEP {\bf 0208} (2002) 036
  [arXiv:hep-th/0206108].
%
\bibitem{Cremades:2002dh}
  D.~Cremades, L.~E.~Ib\'{a}\~{n}ez and F.~Marchesano,
  Nucl.\ Phys.\  B {\bf 643} (2002) 93
  [arXiv:hep-th/0205074];
  arXiv:hep-ph/0212048.
%
\bibitem{Ellis:2002ci}
  J.~R.~Ellis, P.~Kanti and D.~V.~Nanopoulos,
  Nucl.\ Phys.\  B {\bf 647} (2002) 235
  [arXiv:hep-th/0206087];
M.~Axenides, E.~Floratos and C.~Kokorelis,
  JHEP {\bf 0310} (2003) 006
  [arXiv:hep-th/0307255].
%
\bibitem{Abel:2003yx}
  S.~A.~Abel and A.~W.~Owen,
  Nucl.\ Phys.\  B {\bf 682} (2004) 183
  [arXiv:hep-th/0310257];
%
 S.~A.~Abel and B.~W.~Schofield,
  Nucl.\ Phys.\  B {\bf 685} (2004) 150
  [arXiv:hep-th/0311051];
  JHEP {\bf 0506} (2005) 072
  [arXiv:hep-th/0412206];
  S.~A.~Abel, O.~Lebedev and J.~Santiago,
  Nucl.\ Phys.\  B {\bf 696} (2004) 141
  [arXiv:hep-ph/0312157];
  S.~A.~Abel and M.~D.~Goodsell,
  JHEP {\bf 0602} (2006) 049
  [arXiv:hep-th/0512072].
%
%
\bibitem{Cvetic:2001tj}
  M.~Cveti\v{c}, G.~Shiu and A.~M.~Uranga,
  Phys.\ Rev.\ Lett.\  {\bf 87} (2001) 201801
  [arXiv:hep-th/0107143],
  Nucl.\ Phys.\  B {\bf 615} (2001) 3
  [arXiv:hep-th/0107166].
%
\bibitem{Blumenhagen:2002gw}
  R.~Blumenhagen, L.~G\"orlich and T.~Ott,
  JHEP {\bf 0301} (2003) 021
  [arXiv:hep-th/0211059];
  R.~Blumenhagen, D.~L\"ust and S.~Stieberger,
  JHEP {\bf 0307} (2003) 036
  [arXiv:hep-th/0305146].
%
\bibitem{Cvetic:2002pj}
  M.~Cveti\v{c}, I.~Papadimitriou and G.~Shiu,
  Nucl.\ Phys.\  B {\bf 659} (2003) 193
  [Erratum-ibid.\  B {\bf 696} (2004) 298]
  [arXiv:hep-th/0212177];
M.~Cveti\v{c} and I.~Papadimitriou,
  Phys.\ Rev.\  D {\bf 67} (2003) 126006
  [arXiv:hep-th/0303197].
\bibitem{cll}
  M.~Cveti\v{c}, T.~Li and T.~Liu,
  Nucl.\ Phys.\  B {\bf 698} (2004) 163
  [arXiv:hep-th/0403061].
%
\bibitem{Honecker:2003vq}
  G.~Honecker,
  Nucl.\ Phys.\  B {\bf 666} (2003) 175
  [arXiv:hep-th/0303015];
%
  G.~Honecker and T.~Ott,
  Phys.\ Rev.\  D {\bf 70} (2004) 126010
  [Erratum-ibid.\  D {\bf 71} (2005) 069902]
  [arXiv:hep-th/0404055].
%
\bibitem{Larosa:2003mz}
  M.~Larosa and G.~Pradisi,
  Nucl.\ Phys.\  B {\bf 667} (2003) 261
  [arXiv:hep-th/0305224].
%
\bibitem{Dudas:2005jx}
  E.~Dudas and C.~Timirgaziu,
  Nucl.\ Phys.\  B {\bf 716} (2005) 65
  [arXiv:hep-th/0502085].
%
\bibitem{Blumenhagen:2005tn}
  R.~Blumenhagen, M.~Cveti\v{c}, F.~Marchesano and G.~Shiu,
  JHEP {\bf 0503} (2005) 050
  [arXiv:hep-th/0502095];
  C.~M.~Chen, V.~E.~Mayes and D.~V.~Nanopoulos,
  Phys.\ Lett.\  B {\bf 648} (2007) 301
  [arXiv:hep-th/0612087];
  C.~M.~Chen, T.~Li, V.~E.~Mayes and D.~V.~Nanopoulos,
  [arXiv:hep-th/0703280].

\bibitem{Bailin:2006zf}
  D.~Bailin and A.~Love,
  Nucl.\ Phys.\  B {\bf 755} (2006) 79
  [arXiv:hep-th/0603172];
  arXiv:0705.0646 [hep-th].
%
\bibitem{Font:2006na}
  A.~Font, L.~E.~Ib\'{a}\~{n}ez and F.~Marchesano,
  JHEP {\bf 0609} (2006) 080
  [arXiv:hep-th/0607219].
\bibitem{Angelantonj:2002ct}
C.~Angelantonj and A.~Sagnotti,
  Phys.\ Rept.\  {\bf 371} (2002) 1
  [Erratum-ibid.\  {\bf 376} (2003) 339]
  [arXiv:hep-th/0204089];\\
%
  G.~Honecker,
  Mod.\ Phys.\ Lett.\  A {\bf 19} (2004) 1863
  [arXiv:hep-th/0407181];\\
  A.~M.~Uranga,
  Class.\ Quant.\ Grav.\  {\bf 22} (2005) S41 ;  \\
  R.~Blumenhagen, M.~Cveti\v{c}, P.~Langacker and G.~Shiu,
  Ann.\ Rev.\ Nucl.\ Part.\ Sci.\  {\bf 55} (2005) 71
  [arXiv:hep-th/0502005];\\
   R.~Blumenhagen, B.~K\"ors, D.~L\"ust and S.~Stieberger,
  arXiv:hep-th/0610327;\\
  E.~Dudas,
  J.\ Phys.\ Conf.\ Ser.\  {\bf 53} (2006) 567;\\
  F.~Marchesano,
  arXiv:hep-th/0702094.
%
\bibitem{Gmeiner:2005vz}
R.~Blumenhagen, F.~Gmeiner, G.~Honecker, D.~L\"ust and T.~Weigand,
  Nucl.\ Phys.\  B {\bf 713} (2005) 83
  [arXiv:hep-th/0411173];
%
  JHEP {\bf 0601} (2006) 004
  [arXiv:hep-th/0510170];
%
  F.~Gmeiner and M.~Stein,
  Phys.\ Rev.\  D {\bf 73} (2006) 126008
  [arXiv:hep-th/0603019];
  F.~Gmeiner,
  Fortsch.\ Phys.\  {\bf 55} (2007) 111
  [arXiv:hep-th/0608227];
  F.~Gmeiner, D.~L\"ust and M.~Stein,
  arXiv:hep-th/0703011.
%
\bibitem{Dijkstra:2004ym}
  T.~P.~T.~Dijkstra, L.~R.~Huiszoon and A.~N.~Schellekens,
  Phys.\ Lett.\  B {\bf 609} (2005) 408
  [arXiv:hep-th/0403196];
%
  Nucl.\ Phys.\  B {\bf 710} (2005) 3
  [arXiv:hep-th/0411129];
  P.~Anastasopoulos, T.~P.~T.~Dijkstra, E.~Kiritsis and A.~N.~Schellekens,
  Nucl.\ Phys.\  B {\bf 759} (2006) 83
  [arXiv:hep-th/0605226].
%
\bibitem{Dixon:1985jw}
  L.~J.~Dixon, J.~A.~Harvey, C.~Vafa and E.~Witten,
  Nucl.\ Phys.\  B {\bf 261} (1985) 678.
%
  Nucl.\ Phys.\  B {\bf 274} (1986) 285.
%
\bibitem{Erler:1992ki}
  J.~Erler and A.~Klemm,
  Commun.\ Math.\ Phys.\  {\bf 153} (1993) 579
  [arXiv:hep-th/9207111].
%
\bibitem{heterotic1}
  D.~Bailin, A.~Love, W.~A.~Sabra and S.~Thomas,
  Mod.\ Phys.\ Lett.\ A {\bf 10} (1995) 337
  [arXiv:hep-th/9407049],
  Mod.\ Phys.\ Lett.\ A {\bf 9} (1994) 2543
  [arXiv:hep-th/9405031],
  Mod.\ Phys.\ Lett.\ A {\bf 9} (1994) 67
  [arXiv:hep-th/9310008],
%
  Mod.\ Phys.\ Lett.\ A {\bf 9} (1994) 1229
  [arXiv:hep-th/9312122],
  Phys.\ Lett.\ B {\bf 320} (1994) 21
  [arXiv:hep-th/9309133].
%
\bibitem{heterotic2}
  A.~E.~Faraggi, S.~F\"orste and C.~Timirgaziu,
  JHEP {\bf 0608}, 057 (2006)
  [arXiv:hep-th/0605117].
%
\bibitem{heterotic3}
  S.~F\"orste, T.~Kobayashi, H.~Ohki and K.~j.~Takahashi,
  JHEP {\bf 0703} (2007) 011
  [arXiv:hep-th/0612044].
%
\bibitem{Takahashi:2007qc}
  K.~j.~Takahashi,
  JHEP {\bf 0703} (2007) 103
  [arXiv:hep-th/0702025].
%
\bibitem{Ploger:2007iq}
  F.~Pl\"oger, S.~Ramos-S\'{a}nchez, M.~Ratz and P.~K.~S.~Vaudrevange,
  JHEP {\bf 0704} (2007) 063
  [arXiv:hep-th/0702176].
%
%
\bibitem{Blumenhagen:2004di}
  R.~Blumenhagen, J.~P.~Conlon and K.~Suruliz,
  JHEP {\bf 0407} (2004) 022
  [arXiv:hep-th/0404254].
%
\bibitem{Blumenhagen:1999md}
  R.~Blumenhagen, L.~G\"orlich and B.~K\"ors,
  Nucl.\ Phys.\  B {\bf 569} (2000) 209
  [arXiv:hep-th/9908130].
%
\bibitem{Berkooz:1996dw}
  M.~Berkooz and R.~G.~Leigh,
  Nucl.\ Phys.\  B {\bf 483} (1997) 187
  [arXiv:hep-th/9605049].

%
\bibitem{Forste:2000hx}
  S.~F\"orste, G.~Honecker and R.~Schreyer,
  Nucl.\ Phys.\  B {\bf 593} (2001) 127
  [arXiv:hep-th/0008250].
%
\bibitem{Kakushadze:1998eg}
  Z.~Kakushadze,
  Nucl.\ Phys.\  B {\bf 535} (1998) 311
  [arXiv:hep-th/9806008].

\bibitem{Bianchi:1991eu}
  M.~Bianchi, G.~Pradisi and A.~Sagnotti,
  Nucl.\ Phys.\  B {\bf 376} (1992) 365.

\bibitem{Kakushadze:1998bw}
  Z.~Kakushadze, G.~Shiu and S.~H.~H.~Tye,
  Phys.\ Rev.\  D {\bf 58} (1998) 086001
  [arXiv:hep-th/9803141].

\bibitem{Angelantonj:1999jh}
  C.~Angelantonj,
  Nucl.\ Phys.\  B {\bf 566} (2000) 126
  [arXiv:hep-th/9908064].

\bibitem{Angelantonj:1999xf}
  C.~Angelantonj and R.~Blumenhagen,
  Phys.\ Lett.\  B {\bf 473} (2000) 86
  [arXiv:hep-th/9911190].
%
\bibitem{Narain:1986qm}
  K.~S.~Narain, M.~H.~Sarmadi and C.~Vafa,
  Nucl.\ Phys.\  B {\bf 288} (1987) 551.
%
\bibitem{Villadoro:2006ia}
  G.~Villadoro and F.~Zwirner,
  JHEP {\bf 0603} (2006) 087
  [arXiv:hep-th/0602120].
%
\bibitem{Blumenhagen:2000ea}
  R.~Blumenhagen, B.~K\"ors and D.~L\"ust,
  JHEP {\bf 0102} (2001) 030
  [arXiv:hep-th/0012156].
%
\bibitem{Berkooz:1996km}
  M.~Berkooz, M.~R.~Douglas and R.~G.~Leigh,
  Nucl.\ Phys.\  B {\bf 480} (1996) 265
  [arXiv:hep-th/9606139].
%
\bibitem{schel-pheno}
B.\ Schellekens, {\it Talk given at ``String Phenomenology 07'',
  Frascati, Italy, 4--8 June 2007.}
%
\bibitem{Faraggi:1993pr}
  A.~E.~Faraggi,
  Phys.\ Lett.\  B {\bf 326} (1994) 62
  [arXiv:hep-ph/9311312].
%
\bibitem{Donagi:2004ht}
  R.~Donagi and A.~E.~Faraggi,
  Nucl.\ Phys.\  B {\bf 694} (2004) 187
  [arXiv:hep-th/0403272].
%
\bibitem{Blumenhagen:2006ab}
  R.~Blumenhagen and E.~Plauschinn,
  JHEP {\bf 0608} (2006) 031
  [arXiv:hep-th/0604033].
%
 \bibitem{ms}
  F.~Marchesano and G.~Shiu,
  JHEP {\bf 0411} (2004) 041
  [arXiv:hep-th/0409132].
%
\end{thebibliography}
